# Photoacoustic Imaging of Lithium Metal Batteries

Huihui Liu,[1,+] Yibo Zhao,[1,+] Jiasheng Zhou,[1] Ping Li,[1] Shou-Hang Bo,[1,3] and Sung-Liang Chen[1,2,4].

[1]*University of Michigan-Shanghai Jiao Tong University Joint Institute, Shanghai Jiao Tong University, Shanghai 200240, China*

[2]*State Key Laboratory of Advanced Optical Communication Systems and Networks, Shanghai Jiao Tong University, Shanghai 200240, China*

[3]*e-mail:* shouhang.bo@sjtu.edu.cn

[4]*e-mail:* sungliang.chen@sjtu.edu.cn

## Abstract

Increasing demand for high-energy batteries necessitates a revisit of the most energy-dense negative electrode, lithium (Li) metal, which was once abandoned in the 1990s because of safety risks associated with inhomogeneous deposition and stripping (i.e., dendrite growth) during battery cycling. In recent years, to better understand and overcome the Li metal dendrite problem, great efforts have been made to reveal dendrite growth processes using various imaging modalities. However, because of being almost invisible to electrons and X-rays, directly imaging Li metal with the required contrast, spatial and temporal resolutions have always been the challenge. Here, we show that by exploiting photoacoustic effect, microscale-resolution three-dimensional structure of Li protrusions can be clearly visualized within minutes by photoacoustic microscopy (PAM). PAM enables high contrast as well as depth information of Li metal inside the glass fiber separator of a Li/Li liquid electrolyte symmetric cell. Our proof-of-principle experiment introduces a new imaging tool to the Li metal battery community, which could greatly benefit the study of fundamental mechanisms of not only the Li metal dendrite growth in conventional and solid-state batteries, but also sodium and magnesium metals. We believe PAM is a promising in-operando tool for battery diagnostic and prognostic.

## I. INTRODUCTION

Lithium (Li)-ion batteries are ubiquitous in present-day technological applications, ranging from portable devices, electric vehicles to grid-scale stationary energy storage. Li-ion batteries are composed of positive and negative electrodes (two Li reservoirs with different concentrations) which are separated by a polymeric membrane, i.e., a separator. The separator is immersed in Li-ion conducting liquid electrolyte which permits only Li-ion shuffling between the positive and negative electrodes during battery cycling. Simultaneously, electrons flow through the external circuit powering electronic devices.

With increasing demand for higher-energy batteries, it is now a common consensus that graphite anode in Li-ion batteries must be replaced with the most energy-dense Li metal in the next-generation Li metal batteries [1]. Ironically, Li metal anode was the choice when the first rechargeable Li battery was invented in the 1970th [2]. Soon afterwards, however, safety hazards associated with Li metal anode were identified, which halted the development of Li metal batteries. The problem is inhomogeneous deposition (during charge) and stripping (during discharge) of Li metal, which forms protrusions into the separator (commonly referred to as dendrites in literature) instead of smooth deposits, leading to short circuit and even explosion of Li metal batteries [3]. The microstructure of such Li metal protrusions can be mossy, whisker-like or dendritic, and is a complex function of cycling duration, rate, temperature, and electrolyte concentration, to name a few. Solid-state batteries utilizing a mechanically strong and non-flammable solid-state electrolyte have been proposed as a promising solution to suppress dendrite growth of Li metal. However, recent studies have shown that metal dendrites can still grow through the grain boundaries of a solid electrolyte and eventually lead to electrolyte crack and short circuit of the battery [4]. It is therefore of uttermost importance to develop a quantitative understanding of Li metal dendrite growth in conventional liquid electrolyte and current solid-state electrolyte settings, and to identify conditions under which smooth Li deposition of tens of micrometers in thickness can be achieved.

Imaging technologies have been demonstrated as a powerful tool to study dendrite growth [4-15]. For example, scanning and transmission electron microscopy has been widely used to acquire images of Li dendrites with high resolution and high quality [4,9,11-14]. While electron microscopy shows the potential to provide the insight into the formation of dendrites, demanding sample preparation is required. Therefore, it is highly challenging for observation *in situ*, which is essential to track the dynamic evolution of Li metal dendrites during the charge and discharge cycles. Three-dimensional (3D) images of subsurface structures underneath Li metal dendrites were observed with X-ray tomography with resolution on the order of a micrometer [6]. However, more than a thousand images were collected with a series of data processing steps required, which restricts the temporal resolution of this technique, limiting its use to *ex situ* observations only. We further note that Li metal is neither visible to electrons nor to X-rays because Li possesses the third lowest electron density of all chemical elements (just above hydrogen and helium), making the observation of bulk Li metal through electron and X-ray microscopy impossible. Only the surfaces of Li metal which are composed of decomposition products resulting from the side reaction between Li metal and the electrolyte (e.g., LiF and $Li_2CO_3$) can be visualized. Magnetic resonance imaging (MRI) was utilized to non-invasively observe and quantify Li metal microstructures [5,8,10,15]. However, Li is inherently insensitive to MRI (e.g., much less than that of proton [10]), limiting both the spatial and temporal resolution of $^6$Li and $^7$Li MRI [5]. Optical microscopy (OM) offers one possible route to *in situ* imaging of dendrites with high temporal resolutions, yet only two-dimensional (2D) images can be obtained [9]. Finally, because of limited penetration depth, none of these techniques discussed above can directly visualize Li dendrite growth within the separator or solid-state electrolyte membrane, an important area to precisely locate the positions and patterns of short circuits caused by metal protrusions.

Photoacoustic imaging is based on the photoacoustic effect that light absorbed by a material can be converted into heat and the subsequent thermoelastic expansion to generate an acoustic wave. In the past 20 years, it has been extensively explored in the biomedical imaging field to reveal a wide variety of endogenous or exogenous absorbers [16-26]. Since light is highly absorbed by most metals, we anticipate that Li metal can be visualized and quantified by photoacoustic imaging. In this paper, for the first time, we demonstrate that photoacoustic imaging can be exploited to map Li protrusions in Li metal batteries in 3D. A home-built photoacoustic microscopy (PAM) system is used to successfully observe the microstructure of Li protrusions inside the separator of a Li/Li liquid electrolyte symmetric cell, and the imaging depth is expected to be more than ~160 μm. Although we image Li metal *ex situ* to demonstrate the utility of photoacoustic imaging, the obtained results suggest that photoacoustic imaging can potentially be a new tool to realize *in situ*, real-time imaging of Li as well as other metals, such as sodium and magnesium [20,23]. We believe that PAM can also be extended to image Li metal dendrite growth in solid-state batteries. The imaging technique is also cost-effective and easy to operate.

## II. METHODS

In this study, a Li/Li liquid electrolyte symmetric cell, consisting of two Li metal electrodes and a liquid electrolyte layer (Fig. 1(a)), was used to showcase PAM imaging of Li protrusions. The liquid electrolyte layer was fabricated using a glass fiber separator (GFS) (GF/D, Whatman) soaked in 1 M $LiPF_6$ electrolyte solution with 1:1:1 (volumetric ratio) ethylene carbonate (EC): diethyl carbonate (DEC): dimethyl carbonate (DMC) as a solvent. The thickness of the Li electrode was ~240 μm, and that of the GFS was ~2 mm before soaked and <2 mm after soaked. The diameter of the two Li electrodes was ~1.3 cm, and that of the GFS was ~1.5 cm. To facilitate PAM imaging, a flat cross-sectional sidewall surface was prepared by cutting the Li/Li cell through its sandwich stack, as shown in Fig. 1(b). Note that the imaged Li electrode thickness varied from <50 μm to up to ~300 μm (results described later), which was different from the original thickness of the Li electrode (~240 μm, before cutting the Li/Li cell sample (Fig. 1(a))). This is most likely a result of mechanical damage to Li metal during cutting the Li/Li cell sample (Fig. 1(b)). One corner of the Li electrode was further removed to mark the location for imaging after cycling. For the electrochemistry test, the Li/Li cell was first sealed in a stainless steel coin cell case (CR2016, Shenzhen Teensky Technology, Shenzhen, China), as shown in Fig. 1(c), before the galvanostatic (*i.e.*, constant current) charging process of the Li/Li cell. After charging, the Li/Li cell was removed from the stainless steel coin cell. Figure 1(d) illustrates the Li/Li cell before and after charging. The direction of the current determines which of the electrodes that Li is stripped from or is plated onto, as shown in Fig. 1(d). In general, more protrusions will be formed for charging at high areal capacity. In this study, several Li/Li cells were charged at different current densities of 0.1, 0.2, 0.3, 0.5, and 1 $mA/cm^2$, respectively, because current density was considered as a major factor affecting the morphology of deposited Li. For each charging current density, the charging time was fixed at 15 hours, and thus, higher current density resulted in larger amount of Li metal deposition (or stripping), which was also confirmed by PAM imaging (results described later). Figure 1(e) shows Li/Li cell voltages as a function of elapsed time at five different charging current densities. For each charging current density, a representative voltage curve from one Li/Li cell sample was plotted. Before PAM imaging, the Li/Li cell, either with or without the charging process, was sealed in a plastic bag filled with liquid (EC:DMC or silicon oil), which was used to facilitate ultrasound coupling without chemically reacting with or physically dissolving any component of the Li/Li cell.

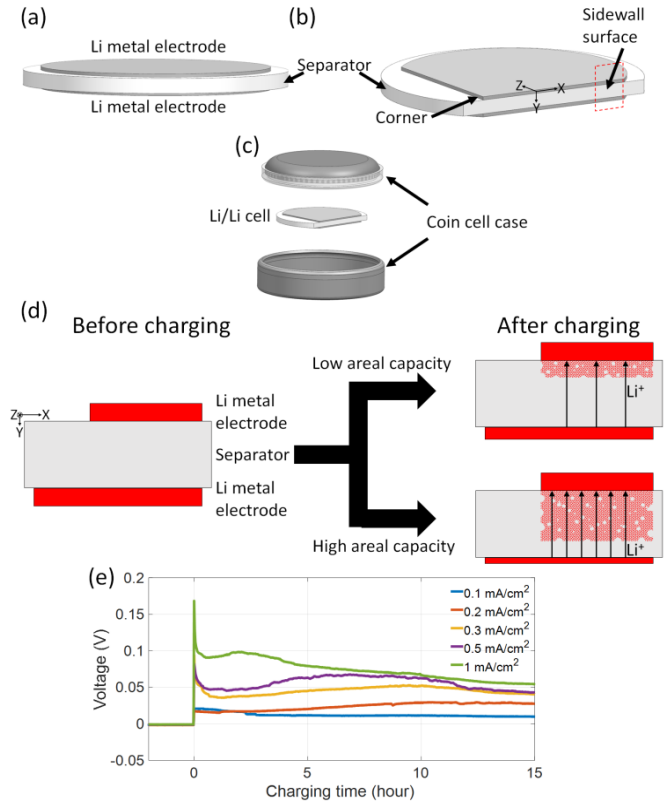

Fig. 1. (a) Schematic of a Li/Li liquid electrolyte symmetric cell. (b) Schematic of the flat cross-sectional sidewall surface of the Li/Li cell with one corner of the Li metal removed as a marker. (c) Schematic of the Li/Li cell sealed in a stainless steel coin cell case. (d) Illustration of the Li/Li cell before and after charging. Current is along the –Y (upward) direction for charging. More protrusions will be formed for charging at high areal capacity. (e) Cell voltages as a function of elapsed time at different charging current densities.

The schematic of the PAM system is shown in Fig. 2(a). A 532 nm pulsed laser (FDSS532-Q4, CryLaS, Germany) was used for photoacoustic imaging. The laser pulse duration was <2 ns, and the repetition rate was 1 kHz. The laser emitted from the laser head was split into two beams by using a beamsplitter (BS025, Thorlabs). The 10% reflected power was detected by a photodiode (DET10A2, Thorlabs) and was used as trigger signals. The 90% transmitted power was attenuated, spatially filtered, and focused by an objective lens (AC254-030, Thorlabs), which was used to excite photoacoustic signals. To facilitate PAM imaging of the cross-sectional sidewall surface of the Li/Li cell, a sample holder was custom made, as shown in Fig. 2(b). By using the sample holder, the Li/Li cell can be stably fixed with its sidewall surface facing upward. A water tank was also custom made, and it was used to facilitate ultrasound coupling. Both the sample holder and the water tank were mounted on a 3D linear motorized stage (M-404, Physik Instrumente [PI], Karlsruhe, Germany), as shown in Fig. 2(a), for scanning during image acquisition. As for detection of photoacoustic waves, a custom-made needle hydrophone (central frequency: 35 MHz) was employed and placed obliquely to the optical axis. Then, the photoacoustic signals were amplified by a preamplifier (ZFL-500LN-BNC+, Mini-Circuits, New York) and sampled by a digitizer (CSE1422, GaGe, Illinois) with sampling rate of 200

MS/s and 14-bit resolution. The data recorded by the digitizer were transferred to a computer for post signal processing and image formation. For post signal processing, a matched filter of 20–60 MHz was adopted to enhance signal-to-noise ratios (SNRs). Note that in this work, the laser energy used was ~86 nJ (per pulse) and no signal averaging was applied unless otherwise specified. The laser energy of ~86 nJ was chosen because it was below the damage threshold of the Li/Li cell used in this work based on our calibration results (see Fig. A1 in Appendix A).

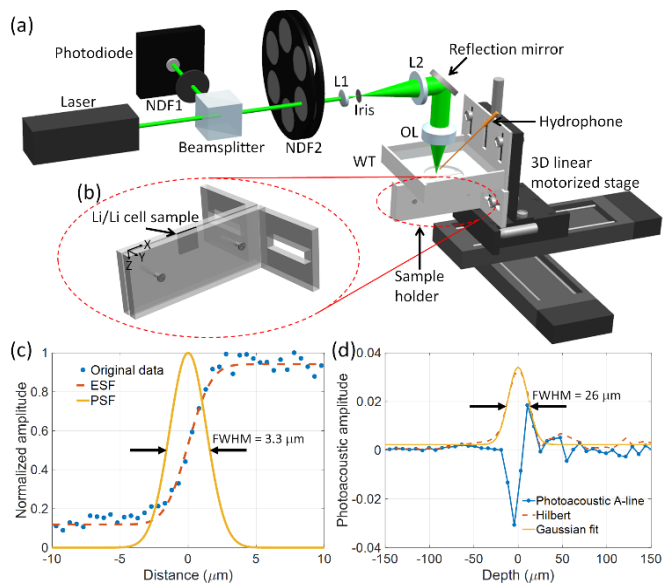

Fig. 2. (a) Schematic of the PAM system. (b) Custom-made sample holder. (c) Calibration of lateral resolution. (d) Measurement of axial resolution. NDF1, neutral density filter 1; NDF2, neutral density filter 2; L1, lens 1; L2, lens 2; OL, objective lens; WT, water tank; PSF, point spread function.

The sharp edge of a razor blade was imaged to calibrate lateral resolution of our PAM system. Figure 2(c) shows the lateral profile of photoacoustic signals with scanning step size of 0.5 μm. The profile was fitted by an edge spread function (ESF) [25]. Then, we took the first derivative of the ESF to obtain a linear spread function and extracted its full width at half maximum (FWHM) as lateral resolution, which was 3.3 μm. Note that higher lateral resolution of PAM can be enabled by using a lens with a higher NA for light focusing. As for measuring axial resolution, a 6-μm-diameter carbon fiber was imaged. Figure 2(d) shows the photoacoustic A-line signal from the carbon fiber and its Hilbert transform (envelope detection). The axial resolution was determined to be 26 μm from the FWHM of the envelope.

## III. RESULTS

To measure the penetration depth of PAM for imaging Li inside the GFS of a Li/Li cell, we devised a sample consisting of three 51-μm-diameter tungsten wires (TWs). Note that the TW (instead of Li) was imaged because of the difficulty in preparing a sample with Li continuously distributed along the depth direction inside the GFS, and because of the similarity in the photoacoustic signal amplitudes from the TW and Li (see Fig. A2 in Appendix B). Sample preparation for the calibration of penetration depth is detailed in Appendix C. The schematic of the sample is shown in Fig. 3(a). The first TW (TW1) was obliquely inserted into the GFS for measurement of penetration depth of PAM. The second TW (TW2) was placed right above the surface of the GFS, which can be used as a reference of the surface of the GFS. The third TW (TW3) was placed above a PET film with thickness of ~150 μm, which was used as a marker for estimation of penetration depth of OM. The PET film was used to ensure the flatness of the top surface of the GFS. The laser energy of ~86 nJ was used and signal averaging of 16 measurements was applied. Figures 3(b) and 3(c) show the PAM maximum amplitude projection (MAP) images of the sample in the XZ and XY planes, respectively. We also checked the depth positions of maximum photoacoustic A-line signal amplitudes along the X direction, as shown in Fig. 3(d). From Figs. 3(b) and 3(d), the penetration depth inside the GFS by PAM was measured to be ~160 μm. Penetration depth can be further enhanced by boosting SNRs, such as using an acoustic detector with higher sensitivity and/or applying more signal averaging. We also measured the penetration depth by using higher laser energy (see Fig. A3 in Appendix D) although it is above the damage threshold mentioned previously. On the other hand, Figs. 3(e) and 3(f) show the OM images of the sample at the foci of TW3 and TW1, respectively. As mentioned above, TW3 was used as a marker. Specifically, we took the OM image of TW3 first, and then took that of TW1 by adjusting the focus of the objective while the lateral position of the sample was kept the same. In this way, we were able to infer the position of TW3 in Fig. 3(f) because Figs. 3(e) and 3(f) are considered to be co-registered in the lateral direction. In Fig. 3(f), the intersection of TW1 and TW3 is denoted as *O*, and the position where TW1 becomes invisible is denoted as *P*. Then, the distance *OP* of ~428 μm can be obtained. Next, by comparing Fig. 3(f) with 3(c), the corresponding positions *O* and *P* in Fig. 3(c) can be determined. Further, by comparing Fig. 3(c) with Figs. 3(b) and 3(d), the corresponding position *P* in Figs. 3(b) and 3(d) can also be labeled, and the depth of TW1 at *P* can be obtained. Finally, the penetration depth inside the GFS by OM was determined to be ~50 μm, which was much shallower than PAM. The results suggest that PAM allows much deeper penetration with high contrast inside the GFS compared with OM, and thus holds promise for 3D visualization of Li inside the GFS.

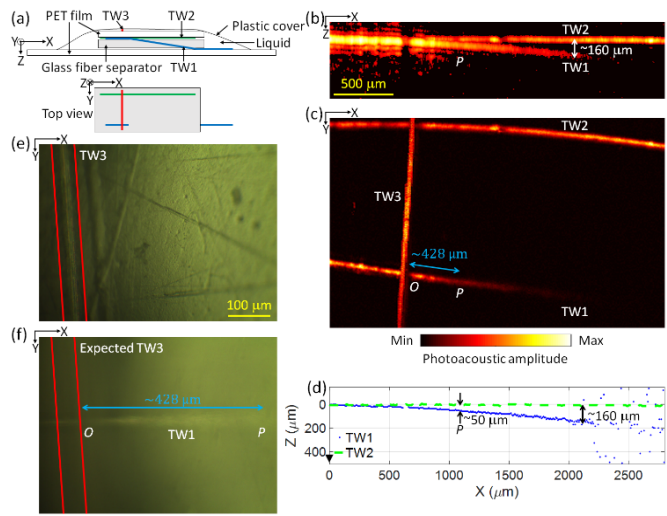

Fig. 3. Calibration of penetration depth. (a) Schematic of the sample consisting of three TWs in side view (upper) and top view (lower). For better illustration, TW1 is plotted in blue, TW2 in green, and TW3 in red. (b,c) PAM MAP images of the sample in the XZ (b) and XY (c) planes. (b) and (c) share the same scale bar in (b). (d) Depth positions of maximum photoacoustic A-line signal amplitudes along the X direction. (e,f) OM images of the sample at the foci of TW3 (e) and TW1 (f). Red lines indicate the positions of TW3 and expected TW3. (e) and (f) share the same scale bar in (e).

Besides, co-registered PAM and OM imaging of the Li-deposited electrode, the side with Li deposition after charging the Li/Li cell, was conducted for further comparison of the two imaging modalities. The Li/Li cell sample was prepared, which was charged under current density of 0.5 mA/cm$^2$ for 15 hours. To facilitate image co-registration of PAM and OM, some makers were made on the top surface of the sample holder (Fig. 4(a)) by using an ink pen. Li was deposited along the +Y direction upon electrochemical charging. Figure 4 shows PAM MAP (XY) and OM images of the Li/Li cell sample. The markers can be clearly seen in Figs. 4(b) and 4(c). Note that Figs 4(c) and 4(e) were taken by 5× and 20× objectives, respectively. As can be seen, by comparing co-registered PAM and OM images in Figs 4(b) and 4(c), PAM enables much higher contrast. Besides, by comparing co-registered PAM and OM images in Figs. 4(d) and 4(e), PAM provides larger depth of focus (DOF). Further, Fig. 4(f) shows the 3D rendering image of Fig. 4(d), demonstrating the 3D imaging capability of PAM. By contrast, OM suffers from low contrast, limited DOF, and no depth information.

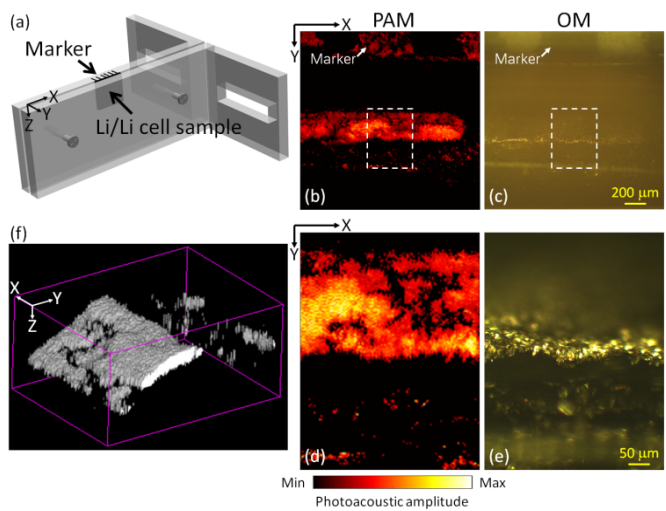

Fig. 4. Comparison of PAM and OM for imaging Li of the Li/Li cell. (a) Schematic of the Li/Li cell sample and markers. (b) PAM MAP (XY) image of the Li/Li cell sample. (c) OM image of the Li/Li cell sample taken by using a 5× objective. (d) Zoom image of the dashed box in (b). (e) OM image of the Li/Li cell sample taken by using a 20× objective, corresponding to the dashed box region in (c). (b) and (c) are co-registered PAM and OM images, and so do (d) and (e). (b) and (c) share the same scale bar in (c), and (d) and (e) share the same scale bar in (e). (f) 3D rendering image of (d). The XYZ orientation is the same as Figs. 1 and 2.

To demonstrate the imaging capability of PAM in visualization of Li protrusions from the Li-deposited electrode towards the GFS of the Li/Li cell, the cross-sectional sidewall surface of one Li/Li cell sample before and after charging at current density of 1 mA/cm$^2$ for 15 hours was imaged. The details of the Li/Li cell sample fabrication, packaging, and charging were mentioned previously. Figure 5 shows the PAM MAP (XY) images at two representative regions around the Li-deposited electrode of the Li/Li cell sample. As can be seen, before charging, a thin layer of the Li metal electrode with relatively uniform thickness was observed. By contrast, after charging, protrusions of Li metal from the Li-deposited electrode towards the GFS can be clearly identified. The fusion image shows the PAM image after charging overlaid with that before charging, providing a better comparison. The results suggest that Li protrusions after charging can be revealed by PAM. As can be observed in Fig. 5, the thickness of the thin layer of the Li electrode before and after charging kept almost the same, which is considered to be plausible. Besides, Li protrusions were concentrated in certain areas of the Li-deposited electrode, demonstrating the inhomogeneous nature of Li deposition. The above-mentioned characteristics were observed in both of the two representative regions in Fig. 5.

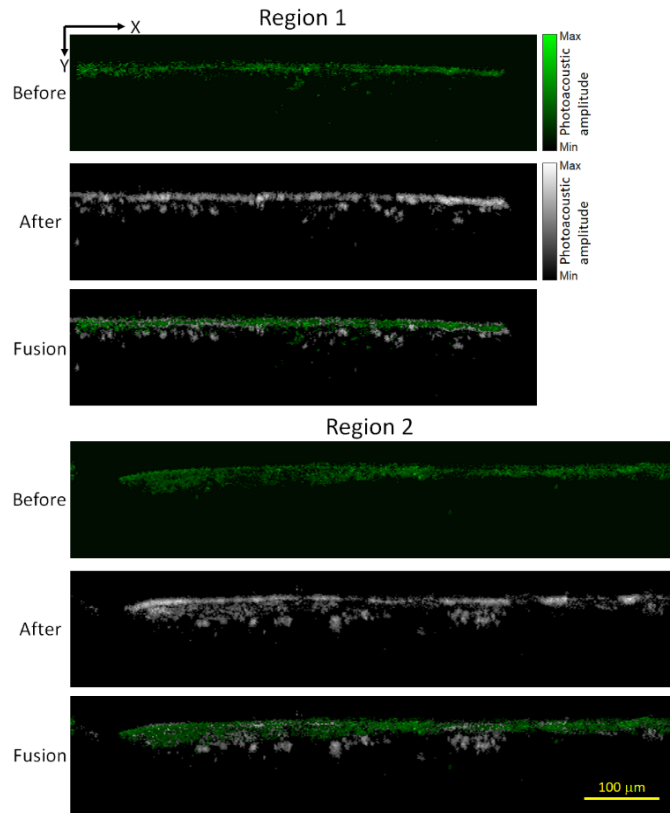

Fig. 5. PAM MAP (XY) images at two representative regions around the Li metal electrode of the Li/Li cell sample before and after charging at current density of 1 mA/cm$^2$ for 15 hours. All images share the same scale bar. The XYZ orientation is the same as Figs. 1 and 2.

Furthermore, because PAM is able to image Li protrusions, as demonstrated above, it is interesting to utilize PAM to study quantitative changes of Li protrusions of the Li/Li cells under different charging current densities. Six Li/Li cell samples were prepared as follows: before charging and after charging under current densities of 0.1, 0.2, 0.3, 0.5, and 1 mA/cm$^2$, respectively, for 15 hours. Figure 6 shows the PAM MAP (XY) images of the six Li/Li cell samples. As can be seen, Li protrusions grew more and more as the charging current density increases. The Li thickness increased from ~0.11 mm for the case of before charging to ~0.54 mm for that of after charging at current density of 1 mA/cm$^2$. Figure 6(b) shows a representative 3D rendering image corresponding to the region labeled by the dashed box in the image of 1 mA/cm$^2$ in Fig. 6(a), demonstrating PAM's ability of 3D examination of Li protrusions inside the GFS. To quantify the Li protrusions under different current densities, the Li ratio, defined as the proportion of the area with Li over the observed area in 2D MAP images, was calculated. As shown in Fig. 6(c), Li ratio increases gradually as increased charging current densities and reaches saturation at current density of 0.5 mA/cm$^2$. For charging current densities <0.5 mA/cm$^2$, the trend agrees with increasing amount of Li deposition as current density (or areal capacity) increases. Above 0.5 mA/cm$^2$, however, no substantial increase of the Li ratio was observed as the current density doubled (i.e., from 0.5 mA/cm$^2$ to 1 mA/cm$^2$), which can be explained as follows. First, local current density over the XZ plane (also see Fig. 1(b)) varies substantially, in particular at high current density, which leads to inhomogeneous deposition of Li (over the XZ plane). Second, instead of the whole XZ plane, only a thin layer (i.e., small thickness along the Z direction) of Li deposition was considered in the calculation of the Li ratio in Fig. 6(c). The above two factors result in the Li ratio not fully aligned with the "average" current density. The results also suggest that PAM has potential for deducing the local current density over the XZ plane from the amount of Li deposition (over the XZ plane) quantified by PAM.

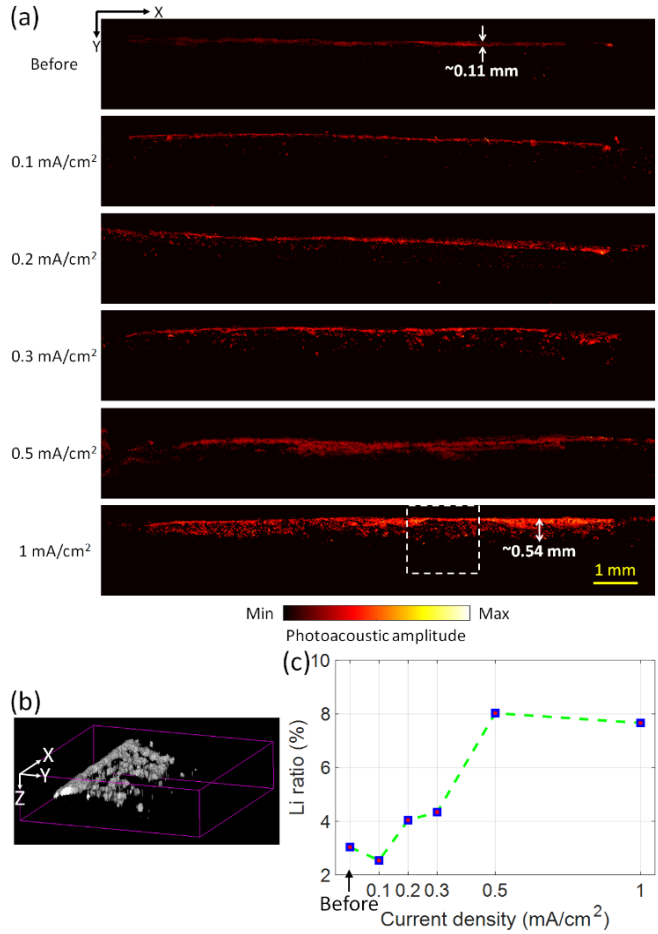

Fig. 6. (a) PAM MAP (XY) images at regions around the Li metal electrode of the six Li/Li cell sample before and after charging at current densities of 0.1, 0.2, 0.3, 0.5, and 1 mA/cm$^2$, respectively, for 15 hours. All images share the same scale bar. (b) 3D rendering image corresponding to the region labeled by the dashed box in the image of 1 mA/cm$^2$ in (a). (c) Quantitative changes of Li ratio. The XYZ orientation is the same as Figs. 1 and 2.

## IV. DISCUSSION AND CONCLUSIONS

Currently, using our PAM system, the image acquisition time for an image consisting of 256 × 256 pixels was ~5 mins. As the aim of this study was to demonstrate the feasibility of this novel PAM approach for imaging Li metal batteries, the imaging speed was not optimized. According to recent studies [20,23], high-speed and even real-time imaging can be realized by using a laser with a high pulse repetition rate and using either a MEMS-mirror scanner or a hexagon-mirror scanner in PAM. Besides, PAM imaging of batteries composed of other metal electrodes such as sodium, magnesium, and zinc would be technically possible and can be explored in future. Development of PAM imaging of Li metal batteries *in situ* would be of great interest for future work.

In summary, PAM was exploited to image Li metal batteries to show potential of PAM as a novel tool to study mechanisms of Li metal dendrite growth. A PAM system with high spatial resolutions was used. Compared with OM, PAM was able to penetrate deeper down to ~160 μm inside the GFS. Further, PAM provided high contrast, large DOF, and depth information for imaging of Li in the Li/Li cell. The 3D rendering image of the Li/Li cell sample acquired by PAM was also demonstrated. Imaging result of one Li/Li cell sample before and after charging demonstrated the PAM ability in observation of Li protrusions. Another imaging result of several Li/Li cells after different charging current densities manifested the potential of PAM in quantitative analysis of Li protrusions. This proof-of-concept study shows that PAM offers a solution to the challenges suffered by existing technologies, such as the prohibitively high cost and demanding sample preparation in electron microscopy. As such, PAM could pave the way to realizing *in situ* observation to facilitate tracking Li metal dendrites during the charge and discharge cycles. There are several advantages and potentials of PAM for imaging Li metal batteries: high resolution (in micrometers), 3D imaging capability, deep penetration into the separator, high contrast from bulk Li metal, and potentials for *in situ* real-time imaging.

**Funding**

National Science Foundation of China (NSFC) (61775134); Shanghai Sailing Program (18YF1411100).

**Appendix A: Calibration of the damage threshold of the Li/Li cell used in this work**

To calibrate the damage threshold of the Li metal of the Li/Li cell under the illumination of 532 nm pulsed laser, several Li foil pieces were imaged by PAM under different laser energy. Since both EC:DMC and silicon oil were used as acoustic coupling media in this work, both liquids were used in this calibration. Four Li metal foil samples sealed in plastic bags were prepared inside an Argon-filled glove box. EC:DMC was used as acoustic coupling media for two samples, and silicon oil for the other two samples. The former two were to be illuminated under laser energy of ~86 nJ and ~185 nJ (per pulse), respectively, and so do the latter two. We first took OM images of the corner of the four Li metal foil samples (the first column in Fig. A1). Then, a small region around the corner of the four samples was imaged by PAM (the dashed boxes in the third column in Fig. A1). After PAM acquisition, OM images of the same corner of the four samples were taken (the second column in Fig. A1). As can be seen in Fig. A1(b) and A1(e) under laser energy of ~86 nJ, OM images after PAM acquisition do not exhibit obvious differences between the PAM regions (referring to the dashed boxes in Fig. A1(c) and A1(f), respectively) and the rest. By contrast, in Fig. A1(h) and A1(k) under laser energy of ~185 nJ, discernible darkened regions corresponding to the PAM regions (referring to the dashed boxes in Fig. A1(i) and A1(l), respectively) for the OM images after PAM acquisition can be clearly identified. The darkened regions were a result of the high-energy pulsed laser which caused the Li metal foil to fall off, thus losing metallic luster. Hence, the damage threshold was determined to be between ~86−185 nJ for the two coupling media, silicon oil and EC:DMC. The laser energy of ~86 nJ was used in this work, as mentioned in the main manuscript. Note that in Fig. A1, by comparing the OM images before and after PAM image acquisition, the latter also shows darkened color randomly, even in the regions without taking PAM. This may be due to the oxidation of Li metal itself during the time elapsed in the experimental process.

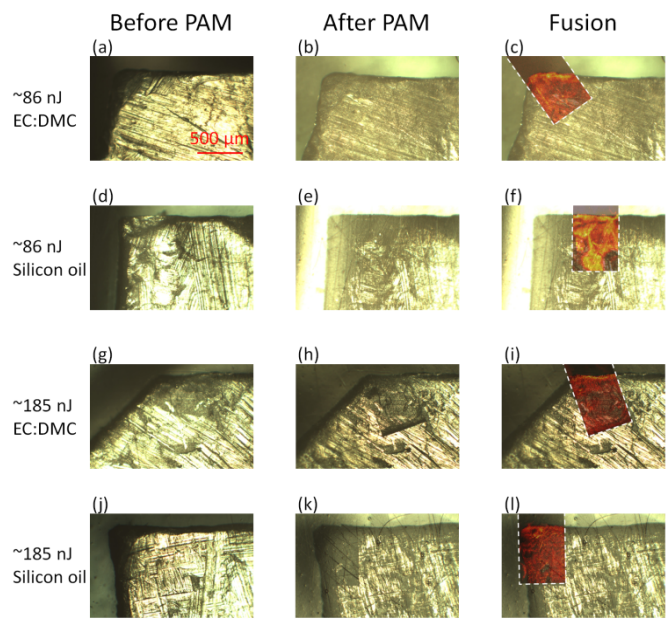

Fig. A1. OM images before and after PAM acquisition under different laser energy and different acoustic coupling media. (a,b,c) Under laser energy of ~86 nJ with EC:DMC as the acoustic coupling medium. (d,e,f) Under laser energy of ~86 nJ with silicon oil as the acoustic coupling medium. (g,h,i) Under laser energy of ~185 nJ with EC:DMC as the acoustic coupling medium. (j,k,l) Under laser energy of ~185 nJ with silicon oil as the acoustic coupling medium. The first column shows the OM images before PAM acquisition, the second column shows the OM images after PAM acquisition, and the third column shows the fusion image of the PAM image (the dashed boxes) and the OM image after PAM acquisition (the second column). All images share the same scale bar.

**Appendix B: Measurement of photoacoustic signal amplitudes from the tungsten wire and Li**

In Fig. 3, the TW (instead of Li) was imaged because of the difficulty in preparing a sample with Li continuously distributed along the depth direction inside the GFS. The results show that the penetration depth by PAM was measured to be ~160 μm. Since Li was expected to be imaged, we here compared the photoacoustic signal amplitudes from the TW and Li. Note that in Fig. 3, the photoacoustic signal of TW1 was measured when it was placed below a PET film with thickness of ~150 μm. Thus, we used the same arrangement (i.e., TW below a PET film) for fair comparison. On the other hand, the photoacoustic signal of Li was measured without a PET film above Li. The laser energy used was ~86 nJ. EC:DMC was used for ultrasound coupling. Figure A2(a) and A2(b) show the PAM MAP (lateral) images of the TW below a PET film and Li, respectively, and Fig. A2(c) shows the top 100 photoacoustic signal amplitudes of them. As can be seen, approximately the photoacoustic signal amplitudes of the TW and Li were similar. The ratio of the average of the top 100 photoacoustic signal amplitudes from the TW to that from Li was also calculated, which was 87%. Since the photoacoustic signal amplitudes from the TW and Li were measured to be similar, the alternative approach to calibrating the penetration depth of PAM was considered to be reasonable. That is, the penetration depth of ~160 μm for Li inside the GFS can be anticipated.

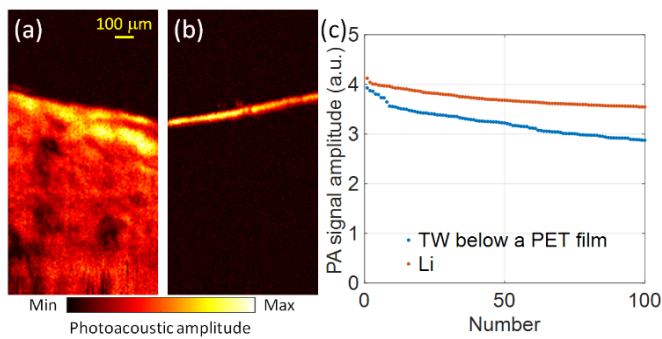

Fig. A2. (a,b) PAM MAP (lateral) images of the TW below a PET film (a) and Li (b). (c) The top 100 photoacoustic signal amplitudes of the TW below a PET film and Li. (a) and (b) share the same scale bar in (a).

**Appendix C: Sample preparation for the calibration of penetration depth**

Figure 3(a) shows the schematic of the sample. First, a large piece of a PET film (~5 cm × 5 cm) was cut, polished by sandpaper, and cleaned by ethanol. It was used as the bottom holder. Then, a piece of the GFS (~1 cm × 1 cm) was prepared. TW1 was obliquely inserted into the GFS. Note that the angle between the inserted TW1 and the surface of the GFS was kept small so that the surface of the GFS at the insertion position of TW1 will not be distorted too much. The two ends of TW1 were fixed on the bottom PET film by waterproof tapes. Then, TW2 was placed above the GFS. As shown in Fig. 3(a), TW1 and TW2 are approximately in parallel along the X direction. Similarly, the two ends of TW1 were fixed on the bottom PET film by waterproof tapes. Another small piece of a PET film (~1 cm × 1 cm) was cut, polished, and cleaned, as mentioned above. As the top cover, the small PET film was placed over the GFS with TW1 and TW2 in between. The top PET film was pressed downward to make sure close contact of TW2 and the surface of the GFS, and then fixed with the bottom PET film by waterproof tapes. TW3 was then placed above the top PET film and fixed by waterproof tapes. Next, a plastic film was used to cover the stack. Three sides of the plastic film were sealed with the bottom PET film by waterproof tapes. Before fully sealing the plastic film, the sample was put inside a glove box. The GFS was filled with EC:DEC:DMC, and then the sample was filled with liquid (EC:DMC) to facilitate ultrasound coupling. Finally, the last side of the plastic film was sealed by waterproof tapes inside the glove box.

**Appendix D: Measurement of penetration depth by using higher laser energy**

In Fig. 3, the penetration depth by PAM was measured to be ~160 μm at excitation laser energy of ~86 nJ. The laser energy above the damage threshold can be used under certain circumstances, for example, only one-time imaging needed. Therefore, we also calibrated the penetration depth by using higher laser energy. The same sample consisting of TWs in Fig. 3 was imaged at laser energy of ~185 nJ and ~357 nJ. Figures A3(a) and A3(b) show the PAM MAP (XZ) images of the sample and corresponding depth positions of maximum photoacoustic A-line signal amplitudes along the X direction at laser energy of ~185 nJ and 357 nJ, respectively. The penetration depth can be deeper than ~220 μm. Interestingly, strong noise was observed with relatively strong photoacoustic signals (left regions in PAM MAP images in Fig. A3), which could be due to the too high laser energy used. The exact reason is under investigation.

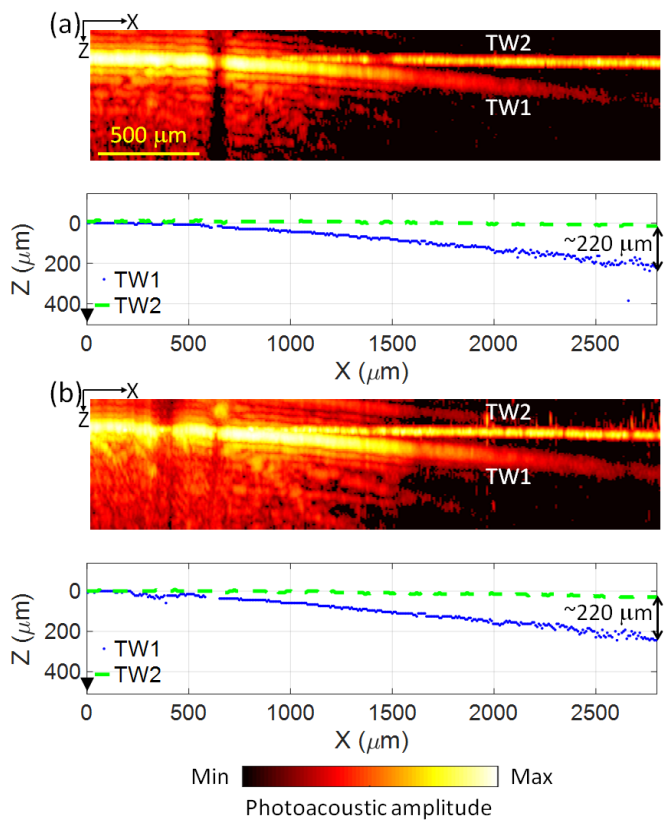

Fig. A3. PAM MAP (XZ) images of the sample consisting of TWs and corresponding depth positions of maximum photoacoustic A-line signal amplitudes along the X direction at laser energy of ~185 nJ (a) and ~357 nJ (b). (a) and (b) share the same scale bar in (a).